\documentclass[10pt, twocolumn,superscriptaddress,showpacs,preprintnumbers]{revtex4}
\usepackage{epsf} 
\usepackage{amsmath}
\usepackage{amssymb}
\usepackage{epsfig}
\usepackage{latexsym}
\usepackage{amsfonts}
\usepackage{dcolumn}
\usepackage{varioref}
\usepackage{ifthen}
\usepackage{graphicx}
\usepackage{amssymb,amsmath,amsthm}
\begin{document}
\parindent 0mm 
\setlength{\parskip}{\baselineskip} 
\thispagestyle{empty}
\pagenumbering{arabic} 
\setcounter{page}{1}
\mbox{ }
\preprint{UCT-TP-281/10}
\preprint{MZ-TH/10-34}
\title	{Charm-quark mass from weighted finite energy QCD sum rules}
\author{S. Bodenstein}
\affiliation{Centre for Theoretical \& Mathematical Physics, University of
Cape Town, Rondebosch 7700, South Africa}
\author{J. Bordes}
\affiliation{Departamento de F\'{\i}sica Te\'{o}rica,
Universitat de Valencia, and Instituto de F\'{\i}sica Corpuscular, Centro
Mixto Universitat de Valencia-CSIC}
\author{C. A. Dominguez}
\affiliation{Centre for Theoretical \& Mathematical Physics, University of
Cape Town, Rondebosch 7700, South Africa}
\affiliation{Department of Physics, Stellenbosch University, Stellenbosch 7600, South Africa}
\author{J. Pe\~{n}arrocha}
\affiliation{Departamento de F\'{\i}sica Te\'{o}rica,
Universitat de Valencia, and Instituto de F\'{\i}sica Corpuscular, Centro
Mixto Universitat de Valencia-CSIC}
\author{K. Schilcher}
\affiliation{Institut f\"{u}r Physik, Johannes Gutenberg-Universit\"{a}t
Staudingerweg 7, D-55099 Mainz, Germany}

\date{\today}
\begin{abstract}
The running charm-quark mass in the $\overline{MS}$ scheme is determined from weighted finite energy QCD sum rules (FESR) involving the vector current correlator. Only the short distance expansion of this correlator is used, together with integration kernels (weights) involving positive powers of $s$, the squared energy. The optimal kernels are found to be a simple {\it pinched} kernel, and polynomials of the Legendre type. The former kernel reduces potential duality violations near the real axis in the complex  s-plane, and the latter allows to extend the analysis to energy regions beyond the end point of the data. These kernels, together with the high energy expansion of the correlator, weigh the experimental and theoretical information differently from e.g. inverse moments FESR. Current, state of the art results for the vector correlator up to four-loop order in perturbative QCD are used in the FESR, together with the latest experimental data. The integration in the complex s-plane is performed using three different methods, fixed order perturbation theory (FOPT), contour improved perturbation theory (CIPT), and a fixed renormalization scale $\mu$ (FMUPT).
The final result is $\bar{m}_c (3\, \mbox{GeV}) = 1008\,\pm\, 26\, \mbox{MeV}$, in a wide region of stability against changes in the integration radius $s_0$ in the complex s-plane.\\
\end{abstract}
\pacs{12.38.Lg, 11.55.Hx, 12.38.Bx, 14.65.Dw}
\maketitle
\noindent
Considerable progress has been made over the years to extract accurate values of the quark masses by confronting QCD with experimental data in the framework of QCD sum rules \cite{REV}. We consider in this paper the case of the charm-quark mass. Following pioneering approaches \cite{SVZ1}, it has been customary to write the Operator Product Expansion (OPE) of the vector correlator with the scale invariant mass $m_c(m_c)$ as the expansion parameter. This is in contrast to the case of light quarks where the square of the four-momentum $Q^2 \equiv -q^2$ is the natural large expansion parameter. If one were to consider the charm quark mass in e.g. the $\overline{MS}$ scheme at a typical scale of $3 \; \mbox{GeV}$, then the fact that $\bar{m}_c(3\;\mbox{GeV}) \simeq 1\;\mbox{GeV}$ should be a matter of some concern regarding this expansion \cite{CADNP}. In any case, the latest determinations based on inverse moment QCD sum rules claim an accuracy at the 1\% level \cite{K}. These inverse moments require QCD knowledge of the vector correlator in the low energy region, around the open charm threshold, as well as in the high energy region.
An alternative approach  involving positive moment sum rules, requiring QCD information only at high energy, and naturally suited to determine $\bar{m}_c(\mu)$ in the $\overline{MS}$ scheme, was proposed some time ago in \cite{KS1}. We  follow this approach here, and make use of  state of the art QCD information up to four-loop level \cite{QCD1}-\cite{QCD12}, together with the  latest experimental data \cite{EXP1}-\cite{EXP4}. We consider Finite Energy Sum Rules (FESR) weighted by two types of integration kernels. The first type is the so called {\it pinched kernel}, which has been shown to minimize potential duality violations close to the real axis in the complex s-plane \cite{PINCH}. The second type is a polynomial (Legendre) kernel tuned to minimize the impact of the energy region where the data is either poor or non-existent. Effectively, these Legendre-type integration kernels, involving several positive powers of the energy, allow for an extension of the analysis beyond the end point of the data. Both types of kernels have been used  successfully in the light quark sector to study the saturation of chiral sum rules \cite{CHSR}, to quantify duality violations \cite{DUAL}, to extract the values of the vacuum condensates in the OPE \cite{COND}, and to determine the  quark condensates \cite{SBARS} and light quark masses \cite{QMASS}.\\
We begin by considering the vector current correlator
\begin{eqnarray}
\Pi_{\mu\nu} (q^2) &=& i \int d^4x \; e^{iqx} \langle 0| T(V_\mu(x) \; V_\nu(0))|0\rangle \nonumber \\ [.3cm]
&=& (q_\mu\; q_\nu - q^2 g_{\mu\nu})\; \Pi(q^2)\;,
\end{eqnarray}
where $V_\mu(x) = \bar{c}(x) \gamma_\mu c(x)$. Invoking Cauchy's theorem in the complex s-plane ($- q^2 \equiv Q^2 \equiv s$) one has
\begin{equation}
\int_{0}^{s_0}
p(s)\; \frac{1}{\pi} Im \;\Pi(s)\;ds = - \frac{1}{2\pi i}
\oint_{C(|s_0|)}
p(s) \;\Pi(s) \;ds \;,
\end{equation}
where $p(s)$ is an arbitrary but analytical integration kernel, and 
\begin{equation}
Im\;\Pi(s) = \frac{1}{12 \pi} \;R_c(s) \;,
\end{equation}
with $R_c(s)$ the standard R-ratio for charm production. The perturbative QCD (PQCD) expression of $\Pi(s)$ at high energies can be written as
\begin{equation}
\Pi(s)|_{PQCD} = \sum_{n=0} \left( \frac{\alpha_s(\mu^2)}{\pi}\right)^n \; \Pi^{(n)}(s) \;,
\end{equation}
where
\begin{equation}
\Pi^{(n)} (s) = \sum_{i=0} \left(\frac{\bar{m}_c^2}{s}\right)^i \; \Pi^{(n)}_i\;,
\end{equation}
and $\bar{m}_c$ stands for the running charm-quark mass in the $\overline{MS}$-scheme.
The complete analytical PQCD result for $\Pi(s)|_{PQCD}$ up to order $\cal{O}$ $[\alpha_s^2 (\bar{m}_c^2/s)^6]$ is given in \cite{QCD1}, and exact results for $\Pi_0^{(3)}$ and $\Pi_1^{(3)}$ are from \cite{QCD2}. The function $\Pi_2^{(3)}$ is known exactly up to a constant \cite{QCD3} which has been estimated using Pad\'{e} approximants in \cite{QCD4}. At five-loop order $\cal{O}$$(\alpha_s^4)$ the full logarithmic terms for $\Pi_0^{(4)}$ are given in \cite{QCD5}, and for $\Pi_1^{(4)}$ in \cite{QCD6}. Since there is incomplete knowledge at this loop-order, we shall use the available information as a measure of the truncation error in PQCD.\\
\begin{figure}
[ht]
\begin{center}
\includegraphics[height=1.9in, width=2.8in]{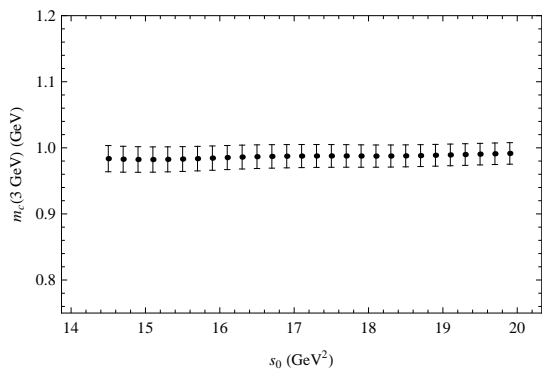}
\caption{The running mass $\bar{m}_c(3\;\mbox{GeV})$ in the  $\overline{MS}$ scheme  for a fixed $\mu=3.0\; \mbox{GeV}$ (FMUPT), as a function of $s_0$ and using the {\it pinched} kernel Eq. (6). Results from FOPT and CIPT lead to similar stability regions. Errors are due to uncertainties in the data, to the values of $\alpha_s$ and the gluon condensate, to PQCD truncation, and to changes in $\mu$ of $\pm\, 50 \%$.}
\end{center}
\end{figure}
The fundamental parameters entering the QCD correlator are the running strong coupling $\alpha_s(\mu^2)$, and the gluon condensate. Regarding the strong coupling, we use the latest comprehensive update analysis at the $\tau$-scale \cite{PICH} which gives $\alpha_s(M_\tau^2)= 0.342 \pm 0.012$, or $\alpha_s(M_Z^2)= 0.1213 \pm 0.0014$. 
This value agrees within errors with a  recent determination \cite{BETH1} from $e^+ e^-$ annihilation data: $\alpha_s(M_Z^2)= 0.1172 \pm 0.0051$.
However, a world average with much smaller errors \cite{BETH2} gives the result $\alpha_s(M_Z^2)= 0.1184 \pm 0.0007$, in agreement with lattice QCD results \cite{LATT}.
We shall include both sets of values of $\alpha_s$ in our determination of $\bar{m}_c$. 
In the non-perturbative sector the leading power correction in the OPE involves the gluon condensate, i.e.
$\left<(\alpha_s/\pi)G^2\right>$. The latest value of the gluon condensate \cite{COND}, extracted from the ALEPH data on $\tau$-decays, is 
$\left<(\alpha_s/\pi)G^2\right>=
(0.046 \pm 0.012)\;\text{GeV}^4$, for $\Lambda_{\text{QCD}}=300\;\text{MeV}$, and
$\left<(\alpha_s/\pi)G^2\right>=
(0.021 \pm 0.006)\;\text{GeV}^4$ for $\Lambda_{\text{QCD}}=350\;\text{MeV}$, 
where $\Lambda_{\text{QCD}}$ stands for the QCD scale in the $\overline{MS}$-scheme. While the gluon condensate is renormalization group invariant, there is an unavoidable dependence on $\alpha_s$ when its value is extracted from data using QCD sum rules. In fact, the sum rules involve the difference between hadronic data integrals and PQCD integrals, with the latter being strongly dependent on the value of $\alpha_s$. In other words, the uncertainty in $\alpha_s$ induces an uncertainty in the condensate.
Extrapolating the above results to include a value $\Lambda_{\text{QCD}}=380\;\text{MeV}$, in line with the latest determination  of $\alpha_s(M_\tau^2)$ \cite{PICH}, leads to
$\left<(\alpha_s/\pi)G^2\right>=(0.01 \pm 0.01)\;\text{GeV}^4$.
This large uncertainty in the value of the gluon condensate will have a very small impact on our results for $\bar{m}_c$. This is because the high energy expansion of the vector correlator, together with integration kernels involving positive powers of $s$, tend to reduce considerably the importance of this term. This is in contrast to the inverse moments method which enhances this contribution with increasing inverse powers of $s$. Something similar happens with the impact of the uncertainty in $\alpha_s$. Results for $\bar{m}_c$ from inverse moments are far more sensitive to $\alpha_s$ than results from sum rules involving the high energy behaviour of the correlator and positive powers of $s$, as with pinched or Legendre integration kernels.\\
Turning to the experimental data, our analysis  follows closely that of \cite{K}, \cite{MC2}. For the first two resonances we use the latest data from the Particle Data Group \cite{PDG}, $M_{J/\psi}= 3.096916 (11)\; \mbox{GeV}$, $\Gamma_{J/\psi} = 5.55 (14) \;\mbox{keV}$, $M_{\psi(2s)}= 3.68609 (4)\; \mbox{GeV}$, $\Gamma_{\psi(2s)} = 2.35 (4) \;\mbox{keV}$. The first two resonances are followed by the open charm region where the contribution from the light quark sector $R_{uds}$ must be subtracted as background from the total R-ratio $R_{tot}$. This we do as in \cite{MC2}. In the region $3.97\;\mbox{GeV} \leq \sqrt{s} \leq 4.26\;\mbox{GeV}$ we only use CLEO data \cite{EXP4} as they have the least error. Regarding the three data sets from BES \cite{EXP1}-\cite{EXP3}, we assume conservatively that the systematic uncertainties are not entirely independent and add them linearly, rather than in quadrature, but we treat them as independent from the CLEO data set \cite{EXP4}, and thus add these in quadrature. The region $s = 25 - 49 \;\mbox{GeV}^2$ has no data, beyond which there is CLEO data up to $s\simeq 110 \;\mbox{GeV}^2$. The latter is fully compatible with PQCD.\\
\begin{figure}
[ht]
\begin{center}
\includegraphics[height=1.9in, width=2.8in]{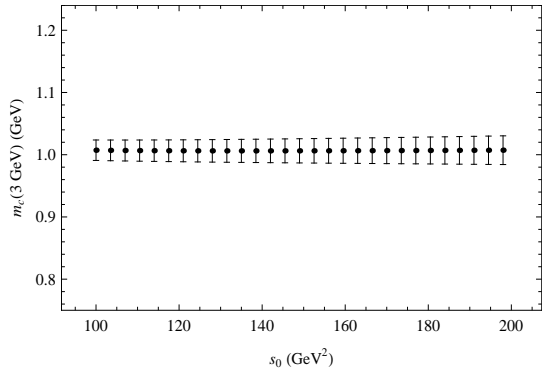}
\caption{The running mass $\bar{m}_c(3\;\mbox{GeV})$ in the  $\overline{MS}$ scheme as a function of $s_0$, using the Legendre-type polynomial $\cal{P}$$_5$ with $s_1 = 24\;\mbox{GeV}^2$. Polynomials of different order lead to equally wide stability regions. The errors are due to uncertainties in the data, the values of $\alpha_s$ and the gluon condensate, and changes in $\mu$ of $\pm\,50 \%$.}
\end{center}
\end{figure}
We discuss next the integration kernels $p(s)$ in Eq.(2), and first introduce the {\it pinched} kernel
\begin{equation}
p(s) = 1 - \frac{s}{s_0} \;.
\end{equation}
As shown in the past \cite{PINCH} - \cite{COND}, this kernel suppresses potential duality violations close to the real axis in the complex s-plane. Clearly, there is a large variety of alternative functional forms vanishing at $s=s_0$. Having considered many of these alternatives, the simple choice above turns out to be the optimal. In any case, higher powers of $s$ will be part of the Legendre-type kernels we discuss next. The purpose of these kernels is to extend the analysis beyond the end point of the data, as achieved e.g. in the analysis of duality violations using the ALEPH $\tau$-decay data \cite{DUAL}. In the present case, there is no data in the interval $25 \;\mbox{GeV}^2 \lesssim s \lesssim 50\; \mbox{GeV}^2$, while the data for $s\gtrsim 50\;\mbox{GeV}^2$ agrees with PQCD. Hence, we introduce a kernel $p(s)$ such that 
\begin{equation}
p(s) = \mathcal{P}_n [x(s)] \;,
\end{equation}
where
\begin{equation}
x(s) = \frac{2 s - (s_0 + s_1)}{s_0 - s_1}\;,
\end{equation}
with $s_0 > s_1$, and $\mathcal{P}_n(x)$ are the standard Legendre polynomials, i.e. $\mathcal{P}_1(x) = x$, $\mathcal{P}_2(x) =(5 x^3 - 3 x)/2$, etc., which satisfy the constraint
\begin{equation}
\int_{s_1}^{s_0} s^k \;\mathcal{P}_n[x(s)]\;ds =0 \;,
\end{equation}
where $s_1 \simeq 24 \; \mbox{GeV}^2$, and $s_0$ varies in the region where there is no data. 
In the hadronic sector, the Legendre-type kernels provide extra weight to the well known resonance region on account of their rapid growth for $s < s_1$.\\
\squeezetable
\begin{table}
\begin{ruledtabular}
\begin{tabular}{cccccccc}
\multicolumn{7}{r}{Uncertainties (MeV)} \\
\cline{3-8}
\noalign{\smallskip}
 Method & $\bar{m}_c(3\,\mbox{GeV})$ & Exp. & $\Delta \alpha_s$  & Trunc. & NP &   Var. & Total               \\
\hline
\noalign{\smallskip}
FOPT\quad &  996 \quad &  9 \quad &\quad  5 \quad &\quad 9 \quad &\quad 1 & \quad 11 \quad &\quad  25 \\
FMUPT\quad &   988 \quad &  9 \quad &\quad  3 \quad &\quad 6 \quad &\quad 1 & \quad 9 \quad &\quad   25 \\
CIPT\quad &   983 \quad &  9 \quad &\quad  1 \quad &\quad 2 \quad &\quad 1 & \quad 16 \quad &\quad   25 \\
\end{tabular}
\caption{\scriptsize{Results for $\bar{m}_c(3 \;\mbox{GeV})$ (in MeV) in the $\overline{MS}$ scheme using the {\it pinched} kernel Eq.(6) in FOPT, CIPT, and in FMUPT using a fixed scale $\mu = 3 \; \mbox{GeV}$. Listed results are at $s_0 = 17 \;\mbox{GeV}^2$, a representative value inside the wide stability region (see Fig. 1). 
The uncertainties are due to the data (Exp.), to $\alpha_s $ ($\Delta \alpha_s$), to truncation of PQCD (Trunc.), to the gluon condensate (NP), and to variation inside the stability region in $s_0$ (Var.). All uncertainties are added in quadrature except for that in $s_0$ which is conservatively added linearly. The FMUPT total error includes an uncertainty in $\mu$ of $\pm\, 50 \%$.}}
\end{ruledtabular}
\end{table}
\begingroup
\squeezetable
\begin{table}
\begin{center}
\begin{ruledtabular}
\begin{tabular}{lccccccc}
\multicolumn{7}{r}{Uncertainties (MeV)} \\
\cline{4-8}
\noalign{\smallskip}
 Kernel & $s_0$ & $\bar{m}_c(3 \, \mbox{GeV})$ & Exp. & $\Delta \alpha_s$ & $\Delta\mu$ &  NP & Total              \\
\hline 
\noalign{\smallskip}
$\mathcal{P}_{3}$&100&  1008 \quad & \quad 24 \quad &\quad  1 \quad &\quad  8 \quad  &\quad  1 \quad &  \quad 25 \\
$\mathcal{P}_{4}$&160  & 1008 \quad &\quad  24 \quad &\quad 1 \quad &\quad  8  \quad&\quad 1  \quad &  \quad 26 \\
$\mathcal{P}_{5}$&200 &  1007 \quad &\quad  22 \quad &\quad 1 \quad &\quad  8  \quad&\quad  2 \quad &   \quad 23 \\
$\mathcal{P}_{6}$&300  & 1008 \quad &\quad  23 \quad &\quad 1 \quad &\quad  9  \quad&\quad  2 \quad &  \quad 25 \\
 \end{tabular} 
\end{ruledtabular}
\caption{\scriptsize{Results for $\bar{m}_c(3 \;\mbox{GeV})$ (in MeV) in the $\overline{MS}$ scheme obtained using  various Legendre polynomial-type kernels, Eqs. (8)-(10), with $s_1=24\; \mbox{GeV}^2$, $\mu = 3\;\mbox{GeV}$, and representative values of $s_0$ in the remarkably wide stability region (see Fig. 2). The uncertainties are due to the data (Exp.), to $\alpha_s $ ($\Delta \alpha_s$), to changes in $\mu$ by $\pm \,50 \%$ ($\Delta \mu$), and to the gluon condensate (NP), which are added in quadrature. The variation in the stability region is negligible, and so is the truncation uncertainty.}}
\end{center}
\end{table}
\endgroup

In Fig. 1 we show the result for $\bar{m}_c(3\; \mbox{GeV})$ as a function of $s_0$ using a fixed scale $\mu = 3\; \mbox{GeV}$ (FMUPT). The errors shown are the result of adding in quadrature the  uncertainties due to experiment, to the values of $\alpha_s$ and  the gluon  condensate, and to the truncation of PQCD which is taken as the difference between the  known four-loop result and the partially known five-loop expression (to zeroth order in $m_c$). This FMUPT total error includes an uncertainty in $\mu$ of $\pm\, 50 \%$ (not present in FOPT or CIPT). This combined error is then conservatively added linearly to the uncertainty due to variation of $s_0$ in the stability region. Results from FOPT and CIPT are similar, and fully compatible within errors. Numerical values are given in Table I. In Fig. 2 we show $\bar{m}_c(3\, \mbox{GeV})$ as a function of $s_0$ for  the kernel $\cal{P}$$_5$ with $s_1= 24\;\mbox{GeV}^2$. Legendre-type polynomials of other order lead to basically the same results, which are very stable in a remarkably wide region. Numerical values are listed in Table II where
the uncertainties are due to the data (Exp.), to $\alpha_s$ ($\Delta \alpha_s$), to changes in $\mu$ by $\pm \,50 \%$ ($\Delta \mu$), and to the gluon condensate (NP). Variation within the stability region gives a negligible uncertainty, and so does the PQCD truncation error. In comparison with values in Table I, one notices the larger impact of experimental uncertainties. This is because $s_0$ is much larger, thus including more data in the hadronic integral.
Since quark mass terms in PQCD are suppressed by inverse powers of $s$, the Legendre-type kernels are expected to improve convergence for increasing energies. We choose as our final result the one from the Legendre-type kernels due to its stability against changes in $s_0$, and use the {\it pinched} kernel values in Table I as confirmation. In this case 
\begin{equation}
\bar{m}_c(3\, \mbox{GeV}) = 1008\, \pm \, 26 \; \mbox{MeV}\;,
\end{equation}
in good agreement within errors with the result from inverse moments QCD sum rules \cite{K}, other recent determinations \cite{MC2}, \cite{SIGNER}, as well as lattice QCD \cite{LATT}. Translated into a scale invariant mass, the above result gives $\bar{m}_c(\bar{m}_c) = 1319 \, \pm \, 26 \; \mbox{MeV}$ for the value used here for the strong coupling, $\alpha_s(M_Z^2)= 0.1213 \pm 0.0014$. Using instead $\alpha_s(M_Z^2) = 0.1189\, \pm \, 0.0020$, as in \cite{K}, changes this mass to $\bar{m}_c(\bar{m}_c) = 1299 \, \pm \, 26 \; \mbox{MeV}$.\\ 
In closing we briefly discuss two convergence issues, (a) the  convergence pattern in $\alpha_s$ of the perturbative QCD integral as a function of $m_c$, and (b) the convergence pattern of the values of $m_c$ resulting from the FESR truncated at different orders in $\alpha_s$. 
In the first case the perturbative contour integral
\begin{equation}
\mathcal{I}(s_0)=-\frac{1}{2\pi i}\oint_{C(|s_0|)}p(s)\Pi(s)\,ds
\end{equation}
is a function of both the mass $\bar{m}$ and the coupling $\alpha_s$. One can investigate the convergence of this integral as a function of $\alpha_s$ for different values of $\bar{m}_c$. Using  a typical integration kernel, e.g. $p(s)=\mathcal{P}_4$ and the representative value $\bar{m}=1\,\text{GeV}	$, we find $\mathcal{I}^{(0)}=1.4176\,\text{GeV}^2$, $\mathcal{I}^{(1)}=1.4563\,\text{GeV}^2$, $\mathcal{I}^{(2)}=1.4541\,\text{GeV}^2$, $\mathcal{I}^{(3)}=1.4512\,\text{GeV}^2$ and $\mathcal{I}^{(4)}=1.4524\,\text{GeV}^2$,  where the upper index in $\mathcal{I}^{(j)}$ indicates the power of $\alpha_s$. If one were to use precisely the same input parameters, but with a higher quark mass, e.g. $\bar{m}=5\,\text{GeV}$, the convergence would expectedly be lost. To restore it one only needs  to increase correspondingly the scale $\mu$. Turning to the second issue, and using again the typical
kernel $\mathcal{P}_4$, one can calculate $\bar{m}_c(3\,\text{GeV})$ truncated at different orders in $\alpha_s$ as a further check on the  convergence. The results are $\bar{m}^{(0)}_c=986\,\text{MeV}$, $\bar{m}^{(1)}_c=1011\, \text{MeV}$, $\bar{m}^{(2)}_c=1010\,\text{MeV}$, $\bar{m}^{(3)}_c=1008\,\text{MeV}$, and $\bar{m}^{(4)}_c=1009\,\text{MeV}$, where the upper index in $\bar{m}^{(j))}$ indicates the power of $\alpha_s$. Considering the difference between $\bar{m}^{(3)}_c$ and $\bar{m}^{(4)}_c$ would give an uncertainty of roughly $1\,\text{MeV}$, which is very much smaller than that from the $50 \%$ variation in the scale $\mu$ considered in our analysis.\\

This work was supported in part by the European FEDER and Spanish MICINN under grant MICINN/FPA 2008-02878, by the Generalitat Valenciana under grant GVPROMETEO 2010-056,  by NRF (South Africa) and by DFG (Germany).
One of us (SB) wishes to thank P. Maier, M. Steinhauser, and C. Sturm for helpful correspondence.
\textsl{}

\end{document}